\documentclass[12pt,draftcls,onecolumn]{IEEEtran}

\usepackage{graphicx}
\usepackage{cite}
\usepackage{times}
\usepackage{amsmath}
\usepackage{amsfonts}
\usepackage{amssymb}

\newfont{\bb}{msbm10 scaled 1100}

\newfont{\bs}{msbm10 scaled 2200}

\newcommand{\Am}{{\bf A}}
\newcommand{\Bm}{{\bf B}}
\newcommand{\Cm}{{\bf C}}

\newcommand{\Wm}{{\bf W}}
\newcommand{\Vm}{{\bf V}}
\newcommand{\Xm}{{\bf X}}
\newcommand{\Ym}{{\bf Y}}
\newcommand{\Zm}{{\bf Z}}

\centerfigcaptionstrue

\def\baselinestretch{1.7}

\begin{document}

\markboth
{\emph{Fertonani and Duman}: Novel Bounds on the Capacity of the Binary Deletion Channel}
{\emph{Fertonani and Duman}: Novel Bounds on the Capacity of the Binary Deletion Channel}

\title{\LARGE{Novel Bounds on the Capacity of the Binary~Deletion~Channel}}

\author
{
\vspace{0.5cm}
{\bf Dario Fertonani and Tolga M. Duman}\\
\vspace{0.2cm}
{\normalsize Department of Electrical Engineering}\\
{\normalsize Arizona State University}\\
{\normalsize Tempe, AZ  85287-5706}\\
{\normalsize E-Mail: dario.fertonani@asu.edu, duman@asu.edu}\\
\vspace{0.5cm}
{
\normalsize
Submission: \today\\
}
}

\def\baselinestretch{1.2}
\maketitle
\thispagestyle{empty}


\begin{abstract}

We present novel bounds on the capacity of the independent and identically distributed binary deletion channel.
Four upper bounds are obtained by providing the transmitter and the receiver with genie-aided information on suitably-defined random processes.
Since some of the proposed bounds involve infinite series, we also introduce provable inequalities that lead to more manageable results.
For most values of the deletion probability, these bounds improve the existing ones and significantly narrow the gap with the available lower bounds.
Exploiting the same auxiliary processes, we also derive, as a by-product, a couple of very simple lower bounds on the channel capacity, which, for low values of the deletion probability, are almost as good as the best existing lower bounds.

\end{abstract}




\section{Introduction\label{s:intro}}

We consider a binary deletion channel where each bit in the input sequence gets deleted, independently of the others, with probability~$d$, while the non-deleted bits are received without errors and in the correct order.
The positions at which the deletions occur are unknown to both the transmitter and the receiver.
Formally, let $\Xm = \{X_n\}_{n=1}^N$~be a sequence of $N$~bits at the input of the channel, let $M$~be the number of received bits, which is a random variable taking values in~$\{0,1,\dots,N\}$ according to the realization of the deletion process, and let $\Ym = \{Y_n\}_{n=1}^M$~be the received sequence.
The capacity per input bit of this channel, generally referred to as independent and identically distributed~(IID) binary deletion channel, is defined as~\cite{Do67}
\begin{equation}
C = \lim_{N\to \infty} \max_{P(\Xm)} \frac{1}{N} I(\Xm;\Ym)
\label{e:C}
\end{equation}
where $P(\Xm)$ is the distribution of the input sequence, and $I(\cdot;\cdot)$ is the average mutual information between two random sequences~\cite{CoTh91}.
The capacity~(\ref{e:C}) is unknown, and only some upper and lower bounds are available in the current literature.

The first lower bound on the capacity of the deletion channel was derived by Gallager in~\cite{Ga61}, where he proved that, \mbox{for~$d\le0.5$}, the capacity of interest is at least equal to that of a binary symmetric channel with bit-flipping probability~$d$.
A number of lower bounds have since been proposed (see \cite{DrMi07}, \cite{DrKi07}, and references therein), among which the best bounds that we are aware of are the ones presented in~\cite{DrMi07} and~\cite{DrKi07}.
In particular, the latter bound outperforms the former when~$d\le0.35$, that is, for all values of~$d$ for which the authors of~\cite{DrKi07} could run the required computations whose execution time grows quickly as $d$~increases.
Throughout the paper, the reference lower bound will thus be the one in~\cite{DrKi07} for~$d\le0.35$ and the one in~\cite{DrMi07} for~$d>0.35$.

Only a few upper bounds have been derived on the capacity of the IID deletion channel.
A simple upper bound is given by the capacity of an IID erasure channel with erasure probability~$d$, since the erasure channel is identical to the deletion channel, except that the receiver additionally knows the positions of the deleted bits~\cite{CoTh91}.
A combinatorial bound proposed by Ullman in~\cite{Ul67}, which was originally derived for particular channels with synchronization errors, had been used for decades as an upper bound for the deletion channel.
However, it is not a true upper bound, and it has been recently found to be violated by provable lower bounds on the capacity of the deletion channel~\cite{DrMi07}.
The reason is due to the fact that Ullman focused on systems with null error probability, while the definition of capacity relies on the weaker condition of error probability that can be made arbitrarily low by increasing the length of the codewords~\cite{CoTh91}.
The only non-trivial upper bound that we are aware of is the one presented in~\cite{DiMiPf07}, which will be adopted here as a reference benchmark.

This paper presents novel upper bounds on the capacity of the IID deletion channel that improve the existing ones for most values of the deletion probability~$d$.
All upper bounds are computed by considering the capacity of some auxiliary channels obtained by providing genie-aided information on suitable random processes related to the deletion process.
In particular, we show that, when such auxiliary random processes are revealed to the transmitter and/or the receiver, we obtain memoryless channels whose capacity can be evaluated by means of the Blahut-Arimoto algorithm~(BAA)~\cite{Bl72,Ar72}, leading to provable upper bounds on the capacity of interest.
Moreover, we show that, based on the introduced auxiliary processes, lower bounds on the capacity of the deletion channel can be derived as well.
The obtained lower bounds, yet close to the ones proposed in \cite{DrMi07} and~\cite{DrKi07} for low values of~$d$, do not improve them, and will only be considered as by-product results.

The paper is organized as follows.
Section~\ref{s:auxiliary} introduces an auxiliary channel based on which we derive three upper bounds on the capacity of the IID deletion channel, which are presented in Sections~\ref{s:ub1}, \ref{s:ub2}, and~\ref{s:ub3}, respectively.
The fourth upper bound, evaluated by exploiting a different auxiliary channel, is introduced in Section~\ref{s:ub4}.
The main contributions in upper bounding the capacity of the deletion channel are summarized an discussed in Section~\ref{s:discussion}.
Finally, Section~\ref{s:lb} introduces a couple of simple lower bounds, while Section~\ref{s:conclusions} gives some concluding remarks.


\section{A Useful Auxiliary Channel\label{s:auxiliary}}

Let $L$ and~$R$ be two natural numbers such that~$R \le L$, and let us define~$D=L-R$.
We consider a channel for which, at each use, the input consists of a sequence of $L$~bits and the output consists of a sequence of $R$~bits.
The input/output relationship characterizing each channel use is the following: $D$~bits are deleted from the $L$~input bits, while the remaining $R$~bits are received without errors and in the correct order.
At each channel use, the deletion pattern, that is, the positions at which the $D$~deletions occur, randomly takes on each of the possible $\binom{L}{D}$~realizations with equal probability, and is unknown to both the transmitter and the receiver.
Also, deletion patterns in different channel uses are independent, so that the channel is memoryless.
As an example, the transition probabilities characterizing the use of the channel are reported in Table~\ref{tab:aux_3_2}, for the case $L=3$ and~$R=2$.
$\Am$ and $\Bm$~denote the input sequence and the output sequence, respectively, while $P(\cdot|\cdot)$~denotes conditional probability.

\begin{table}
\centering
\begin{tabular}{||c||c|c|c|c||}
\hline
\hline
& \multicolumn{4}{|c||}{$P(\Bm|\Am)$} \\
\hline
$\Am$ & $\Bm=00$ & $\Bm=01$ & $\Bm=10$ & $\Bm=11$ \\
\hline
\hline
$000$ & $1$ & $0$ & $0$ & $0$ \\
\hline
$001$ & $1/3$ & $2/3$ & $0$ & $0$ \\
\hline
$010$ & $1/3$ & $1/3$ & $1/3$ & $0$ \\
\hline
$011$ & $0$ & $2/3$ & $0$ & $1/3$ \\
\hline
$100$ & $1/3$ & $0$ & $2/3$ & $0$ \\
\hline
$101$ & $0$ & $1/3$ & $1/3$ & $1/3$ \\
\hline
$110$ & $0$ & $0$ & $2/3$ & $1/3$ \\
\hline
$111$ & $0$ & $0$ & $0$ & $1$ \\
\hline
\hline
\end{tabular}
\caption{Transition probabilities for the auxiliary channel.}
\label{tab:aux_3_2}
\end{table}

The capacity per use of the considered auxiliary channel is defined as
\begin{equation}
f(L,R) = \max_{P(\Am)} I(\Am;\Bm) \;,
\label{e:cap_fLR}
\end{equation}
where $P(\Am)$ is the distribution of the input sequence.
Since each channel output is a sequence of $R$~bits, the following upper bound holds
\begin{equation}
f(L,R) \le R \;.
\label{e:ub_fLR}
\end{equation}
In some particular cases, it can be shown that $f(L,R)$ achieves the upper bound~(\ref{e:ub_fLR}).
These cases are listed and briefly discussed in the following.
\begin{itemize}
\item $f(L,0) = 0$. All input bits are deleted and no information can be delivered.

\item $f(L,1) = 1$. A capacity-achieving scheme consists of transmitting, at each channel use, \mbox{either a} sequence of $L$ zeros or a sequence of $L$ ones, with equal probability and independently of the previous/future transmissions. In this case, for each channel use, the only received bit fully determines the input sequence, irrespectively of the deletion pattern. Formally, adopting the standard notation for the entropy and the conditional entropy~\cite{CoTh91}, we get
\begin{equation}
I(\Am;\Bm) = H(\Am) - H(\Am|\Bm) = H(\Am) = 1
\nonumber
\end{equation}
which achieves the upper bound~(\ref{e:ub_fLR}).

\item $f(L,L) = L$. Since all transmitted bits are correctly received, the capacity is equal to $L$~bits per channel use, which is achieved by independent and uniformly distributed~(IUD) input bits.
\end{itemize}
When~$R\notin \{0,1,L\}$, we could not find a closed-form expression of the capacity~$f(L,R)$.
On the other hand, since the auxiliary channel is memoryless and has finite input/output alphabets, its capacity can be numerically evaluated by means of the BAA~\cite{Bl72,Ar72}.
To run the BAA, we only need the transition probabilities characterizing the channel, as those reported in Table~\ref{tab:aux_3_2}.
Hence, in principle, we can compute the capacity~$f(L,R)$ based on similar tables, for all desired values of $L$ and~$R$.
Unfortunately, the implementation of the BAA becomes computationally infeasible for large values of~$L$ --- for example, $L=17$ is the largest value that we were able to manage for all possible values of~$R$, while $L=22$ is the largest value that we were able to manage for~$R=L-1$, which will be shown later to be a case of particular interest.
Some values of~$f(L,R)$ are reported in Table~\ref{tab:fLR}, where the results obtained by means of the BAA have a two-digit precision after the decimal point, and are rounded up to the next hundredth since, rigorously, the BAA can underestimate the true capacity if a finite number of iterations are performed~\cite{Bl72,Ar72}.

\begin{table*}
\centering
\begin{tabular}{||c||c|c|c|c|c|c|c|c||}
\hline
\hline
 & $R=0$ & $R=1$ & $R=2$ & $R=3$ & $R=4$ & $R=5$ & $R=6$ & $R=7$ \\
\hline
\hline
$L=0$ & 0 & & & & & & & \\
\hline
$L=1$ & 0 & 1 & & & & & & \\
\hline
$L=2$ & 0 & 1 & 2 & & & & & \\
\hline
$L=3$ & 0 & 1 & 1.48 & 3 & & & & \\
\hline
$L=4$ & 0 & 1 & 1.35 & 2.18 & 4 & & & \\
\hline
$L=5$ & 0 & 1 & 1.30 & 1.88 & 2.87 & 5 & & \\
\hline
$L=6$ & 0 & 1 & 1.28 & 1.77 & 2.43 & 3.62 & 6 & \\
\hline
$L=7$ & 0 & 1 & 1.26 & 1.71 & 2.23 & 3.04 & 4.41 & 7 \\
\hline
\hline
\end{tabular}
\caption{Capacity $f(L,R)$.}
\label{tab:fLR}
\end{table*}

In the following, we introduce several lemmas that will be used in the remaining sections to manipulate the capacity of the auxiliary channel when running the BAA seems impossible.
Before providing the lemmas, we define
\begin{equation}
\tilde{f}(L,D) = f(L,L-D) \;,
\label{e:cap_tfLD}
\end{equation}
so that we can index the capacity of the auxiliary channel either by the number of received bits, using~$f(\cdot,\cdot)$, or by the number of deleted bits, using~$\tilde{f}(\cdot,\cdot)$.
The following definitions will also be useful in the remaining sections:
\begin{eqnarray}
\alpha(L,R) &=& R - f(L,R) \;,
\label{e:alphaLR} \\
\tilde{\alpha}(L,D) &=& \alpha(L,L-D) = L-D - \tilde{f}(L,D) \;.
\label{e:talphaLD}
\end{eqnarray}
Note that the coefficients $\alpha(\cdot,\cdot)$ and~$\tilde{\alpha}(\cdot,\cdot)$ cannot be negative due to~(\ref{e:ub_fLR}).

\textit{Lemma 1:} For all values of $L$ and $R$, the following holds
\begin{equation}
f(L+1,R) \le f(L,R) \;.
\label{e:lemma1}
\end{equation}
The proof is based on the fact that, when additional information is provided to the transmitter, the capacity of a system cannot decrease~\cite{CoTh91}.
In particular, the capacity~$f(L+1,R)$ cannot decrease if, at each channel use, the transmitter knows one of the positions at which the $L+1-R$~deletions occur.
Clearly, the bit transmitted in that position is irrelevant.
Moreover, if the revealed position is chosen according to a uniform distribution on the $L+1-R$~possible values, the system is characterized by $L$~effective input bits, $R$~output bits, and IUD deletion patterns, that is, by definition, a system with capacity~$f(L,R)$.
Hence, the lemma is proved.

\textit{Lemma 2:} For all values of $L$ and $R$, the following holds
\begin{equation}
\textrm{if} \quad \hat L > L \quad \textrm{then} \quad \alpha(\hat L,R) \ge \alpha(L,R) \;.
\label{e:lemma2}
\end{equation}
The proof that $\alpha(L+1,R) \ge \alpha(L,R)$ is simply derived from (\ref{e:alphaLR}) and~(\ref{e:lemma1}).
The remainder of the lemma can then be proved by induction.

\textit{Lemma 3:} For all values of $L$ and all positive values of~$D$, the following holds
\begin{equation}
\tilde{f}(L+1,D) \le \tilde{f}(L,D-1)\frac{D}{L+1} + \left[\tilde{f}(L,D)+1\right] \left(1-\frac{D}{L+1}\right) \;.
\label{e:lemma3}
\end{equation}
The proof is based on the fact that, when additional information is provided to both the transmitter and the receiver, the capacity of a system cannot decrease~\cite{CoTh91}.
In particular, we consider the information on the binary event ``the last bit of the $L+1$~transmitted bits is deleted'', which occurs with probability $D/(L+1)$.
When the event occurs, the last transmitted bit is irrelevant and the system is characterized by $L$~effective input bits and $D-1$~deletions on IUD positions, that is, the system has capacity~$\tilde{f}(L,D-1)$.
When the event does not occur, the last transmitted bit can be safely sent uncoded, while, for the first $L$ transmitted bits, the systems is characterized by $L$~effective input bits and $D$~deletions on IUD positions, that is, the system has capacity~$\tilde{f}(L,D)$.
Hence, the lemma is proved.

\textit{Lemma 4:} For all values of $L$ and all values of~$D$, the following holds
\begin{equation}
\tilde{\alpha}(L+1,D) \ge \tilde{\alpha}(L,D) \left(1-\frac{D}{L+1}\right) \;.
\label{e:lemma4}
\end{equation}
The lemma is proved after straightforward manipulations of~(\ref{e:lemma3}) based on (\ref{e:ub_fLR}) and~(\ref{e:talphaLD}).

The lemmas provided hereafter focus on a particular case, that is, the occurrence of exactly one deletion.
The reader may skip them without affecting the arguments exploited in Sections~\ref{s:ub1}, \ref{s:ub2}, \ref{s:ub3}, and~\ref{s:ub4}.
The interest for this case will become evident in Section~\ref{s:discussion}.

\textit{Lemma 5:} For all values of $L$, the following holds
\begin{equation}
\tilde{f}(nL,1) \le \tilde{f}(L,1) + (n-1)L , \qquad \forall n>0 \;.
\label{e:lemma5}
\end{equation}
Let us partition the input sequence of $nL$~bits into $n$~subsequences of $L$~consecutive bits, and let us assume that both the transmitter and the receiver knows in which of the subsequences the deletion occurs.
By definition, this subsequence has capacity~$\tilde{f}(L,1)$, while the remaining $n-1$~subsequences have capacity~$L$.
Hence, since the capacity~$\tilde{f}(nL,1)$ cannot exceed that of the described genie-aided system, the lemma is proved.

\textit{Lemma 6:} For all values of $L$, the following holds
\begin{equation}
\tilde{\alpha}(nL,1) \ge \tilde{\alpha}(L,1) , \qquad \forall n>0 \;.
\label{e:lemma6}
\end{equation}
The lemma directly follows from~(\ref{e:lemma5}) by definition~(\ref{e:talphaLD}).

\textit{Lemma 7:} For all values of $L$, the following holds
\begin{equation}
\tilde{f}(L+1,1) \ge \tilde{f}(L,1) + 1 - \frac{1}{L+1} - h \left( \frac{1}{L+1} \right)
\label{e:lemma7}
\end{equation}
where~$h(\cdot)$ is the binary entropy function~\cite{CoTh91}.\\
To prove the lemma, we first notice that the equation
\begin{equation}
I(\Am;\Bm) = I(\Am;\Bm,\Cm) - I(\Am;\Cm|\Bm)
\nonumber
\end{equation}
holds irrespectively of the definition of the random processes $\Am$, $\Bm$, and~$\Cm$~\cite{CoTh91}.
Moreover, since $I(\Am;\Cm|\Bm)$~cannot be larger than the entropy~$H(\Cm)$ of the process~$\Cm$, we can write
\begin{equation}
I(\Am;\Bm) \ge I(\Am;\Bm,\Cm) - H(\Cm) \;.
\label{e:lemma7_proof}
\end{equation}
In particular, let $\Am$ and $\Bm$~be, respectively, the input sequence and the output sequence of the auxiliary channel considered in this section, when the input sequence includes $L+1$~bits and exactly one deletion occurs.
Also, let $\Cm$~be the binary event ``the last bit of the $L+1$~transmitted bits is deleted'', whose entropy is~$H(\Cm)=h(1/(L+1))$.
Under these definitions, the inequality
\begin{equation}
\tilde{f}(L+1,1) \ge \max_{P(\Am)} I(\Am;\Bm,\Cm) - h \left( \frac{1}{L+1} \right)
\label{e:lemma7_proof_2}
\end{equation}
follows from~(\ref{e:lemma7_proof}).
Note that the first term at the right-hand side of~(\ref{e:lemma7_proof_2}) is the capacity of a channel identical to the considered one, when the receiver is provided with side information on the event~$\Cm$, while the transmitter is not.
According to the data-processing inequality~\cite{CoTh91}, the capacity of this genie-aided system does not increase if, when the event~$\Cm$ occurs, the receiver deletes one of the received bits, selected with equal probability over the $L$~received bits.
In this case, the channel consists of two independent subchannels: the former is characterized by $L$~input bits and one deletion on IUD positions, and thus has capacity~$\tilde{f}(L,1)$, while the latter is an erasure channel with erasure probability~$1/(L+1)$, and thus has capacity~$1-1/(L+1)$.
Hence, we can write
\begin{equation}
\max_{P(\Am)} I(\Am;\Bm,\Cm) \ge \tilde{f}(L,1) + 1 - \frac{1}{L+1}
\nonumber
\end{equation}
which, combined with~(\ref{e:lemma7_proof_2}), proves the lemma.

\textit{Lemma 8:} The following holds
\begin{equation}
\lim_{L\to\infty} \frac{\tilde{\alpha}(L+1,1)}{\tilde{\alpha}(L,1)} = 1 \;.
\label{e:lemma8}
\end{equation}
To prove the lemma, we first notice that the inequality
\begin{equation}
\tilde{\alpha}(L+1,1) \le \tilde{\alpha}(L,1) + \frac{1}{L+1} + h \left( \frac{1}{L+1} \right)
\label{e:lemma8_proof}
\end{equation}
directly follows from~(\ref{e:lemma7}) by definition~(\ref{e:talphaLD}).
Then, according to (\ref{e:lemma4}) and~(\ref{e:lemma8_proof}), we can write
\begin{equation}
1- \frac{1}{L+1} \le \frac{\tilde{\alpha}(L+1,1)}{\tilde{\alpha}(L,1)} \le  1 + \frac{1}{\tilde{\alpha}(L,1)(L+1)} + \frac{1}{\tilde{\alpha}(L,1)} h \left( \frac{1}{L+1} \right)
\nonumber
\end{equation}
which proves the lemma since both sides tend to~one as $L$~tends to infinity.

\textit{Lemma 9:} The following holds
\begin{equation}
\lim_{L\to\infty} \left[\tilde{f}(L+1,1)-\tilde{f}(L,1)\right] = 1 \;.
\label{e:lemma9}
\end{equation}
To prove the lemma, we first notice that the inequalities
\begin{equation}
1 - \frac{1}{L+1} - h \left( \frac{1}{L+1} \right) \le \left[\tilde{f}(L+1,1)-\tilde{f}(L,1)\right] \le  1 + \frac{L-1-\tilde{f}(L,1)}{L+1}
\label{e:lemma9_proof}
\end{equation}
follow from (\ref{e:lemma3}) and~(\ref{e:lemma7}) after simple manipulations.
The left-hand side in~(\ref{e:lemma9_proof}) clearly tends to~one as $L$~tends to infinity.
Then, we notice that the limit
\begin{equation}
\lim_{L\to\infty} \frac{\tilde{f}(L,1)}{L} = 1
\nonumber
\end{equation}
follows from the fact that the binary channel with one deletion tends to the binary identity channel, whose capacity per input bit is one, as the length of the input sequence tends to infinity. 
Hence, the right-hand side in~(\ref{e:lemma9_proof}) tends to~one as $L$~tends to infinity, and the lemma is proved.
Note that~(\ref{e:lemma9}) implies~(\ref{e:lemma8}), but is stronger.

\section{The First Upper Bound\label{s:ub1}}

\begin{figure}
\centering
\includegraphics[width=12cm]{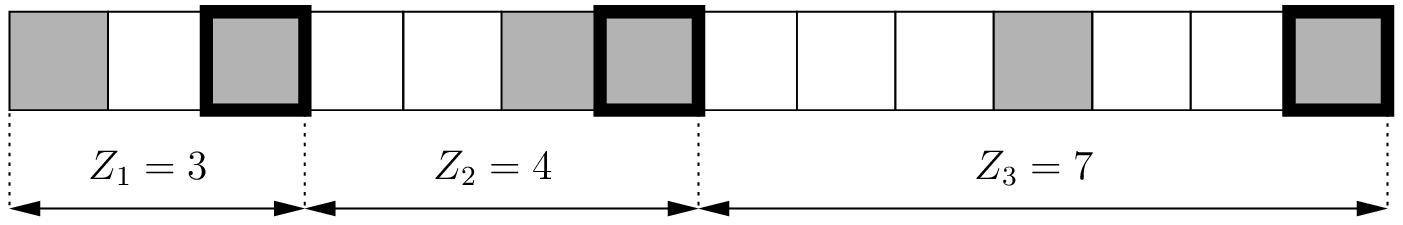}
\caption{A possible realization of the process~$\Zm$, when $D=1$ and~$S=3$. Each white square represents a transmitted bit that is correctly received, while each gray square represents a transmitted bit that is deleted. The positions of the bold-faced bits define the process~$\Zm$.}
\label{fig:Z}
\end{figure}

In this section, we derive an upper bound on~$C$ by providing side information on a random process~$\Zm$, defined in the following.
Let $D$ be a non-negative integer parameter and let us assume that the total number of deleted bits is a multiple of~$D+1$, so that $S=(N-M)/(D+1)$ is an integer --- this assumption does not affect the capacity evaluation, where the limit $N\to \infty$ is to be considered.
We define $\Zm=\{Z_i\}_{i=1}^S$ such that $Z_1$ is equal to the position in the transmitted sequence of the $[D+1]$-th~deleted bit and, for each value of~$i$ in $\{2,3,\dots,S\}$, $Z_i$ is equal to the difference between the position in the transmitted sequence of the $[(D+1)i]$-th~deleted bit and that of the $[(D+1)(i-1)]$-th~deleted bit.
An example is depicted in Fig.~\ref{fig:Z} and discussed in the related caption.
Given the assumption of IID deletions, the process~$\Zm$ is IID too, and each element of~$\Zm$ takes on the value $L+1$ with probability
\begin{equation}
P(Z_i=L+1) = \binom{L}{D} d^{D+1} (1-d)^{L-D}
\label{e:ub1_pascal_pre}
\end{equation}
according to the Pascal distribution~\cite{Pa91}, for all values of $L$ such that~$L\ge D$.
To point out various similarities between the bounds presented in this paper, it is useful to define, for $L\ge R\ge0$ and $L\ge D\ge0$, the terms
\begin{eqnarray}
p(L,R) &=& \binom{L}{R} d^{L-R} (1-d)^R
\label{e:pLR}
\\
\tilde{p}(L,D) &=& p(L,L-D) = \binom{L}{D} d^{D} (1-d)^{L-D}
\label{e:tpLR}
\end{eqnarray}
so that we get
\begin{equation}
P(Z_i=L+1) = d \cdot \tilde{p}(L,D) \;.
\label{e:ub1_pascal}
\end{equation}

The realizations of the process~$\Zm$ are actually unknown to both the transmitter and the receiver.
Hence, an upper bound on the capacity of the deletion channel can be obtained by providing them with genie-aided information on~$\Zm$.
We will refer to the capacity per input bit of this genie-aided system as~$C_1$.
With this side information, we have $S$~blocks that do not interfere with each other, where the $i$-th block has $Z_i$ input bits,  $D+1$ of which get deleted.
The last input bit of each block is irrelevant, since both the transmitter and the receiver know that it gets deleted.
The $i$-th block is thus characterized by $Z_i-1$ effective input bits and $D$ deletions on IUD positions, so that the related capacity is~$\tilde{f}(Z_i-1,D)$, as defined in Section~\ref{s:auxiliary}.
Hence, defining the expectation operator~$E[\cdot]$ and considering that
\begin{equation}
\lim_{N\to \infty} \frac{N}{S} = E[Z_i]
\nonumber
\end{equation}
by the law of large numbers~\cite{Pa91}, we get
\begin{eqnarray}
C_1 &=& \lim_{N\to \infty} \frac{1}{N} \sum_{i=1}^S \tilde{f}(Z_i-1,D)
\nonumber \\
&=& \frac{1}{E[Z_i]} \lim_{S\to \infty} \frac{1}{S} \sum_{i=1}^S \tilde{f}(Z_i-1,D)
\nonumber \\
&=& \frac{1}{E[Z_i]} E \left[ \tilde{f}(Z_i-1,D)  \right]
\nonumber
\end{eqnarray}
where the last equality follows from the law of large numbers.
Finally, by exploiting the properties of the Pascal distribution~\cite{Pa91}, the upper bound yields
\begin{equation}
C_1 = \frac{d^2}{D+1} \sum_{L=D}^\infty \tilde{f}(L,D) \tilde{p}(L,D)
\nonumber
\end{equation}
which can be also written as
\begin{eqnarray}
C_1 &=& \underbrace{\frac{d^2}{D+1} \sum_{L=D}^\infty \left[L-D\right] \tilde{p}(L,D) }_{1-d} - \frac{d^2}{D+1} \sum_{L=D}^\infty \left[L-D-\tilde{f}(L,D)\right] \tilde{p}(L,D)
\nonumber\\
&=& 1-d - \frac{d^2}{D+1} \sum_{L=D}^\infty \tilde{\alpha}(L,D) \tilde{p}(L,D) \;.
\label{e:C_1}
\end{eqnarray}
Since the coefficients $\tilde{\alpha}(\cdot,\cdot)$ cannot be negative, the bound~(\ref{e:C_1}) is at least as good as the trivial bound $1-d$.
In particular, by combining Lemma~4 with the available outcomes of the BAA, it can be proved that the bound~(\ref{e:C_1}) equals~$1-d$ when~$D=0$, otherwise it is strictly better.

\begin{figure}
\centering
\includegraphics[width=12cm]{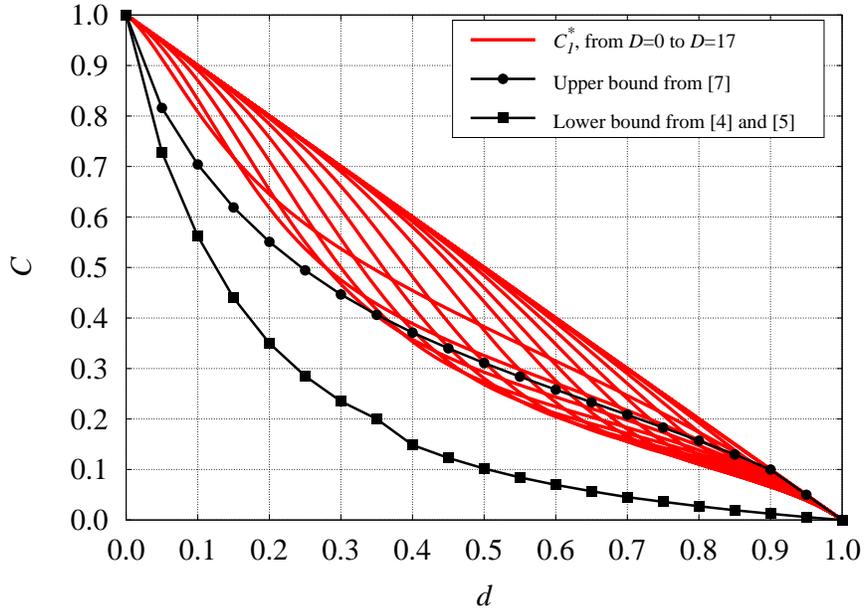}
\caption{Different bounds on the capacity of the deletion channel.}
\label{fig:ub1}
\end{figure}

Unless~$D=0$, it seems infeasible to evaluate the coefficients $\tilde{\alpha}(L,D)$ for all values of~$L$ required in~(\ref{e:C_1}).
Let us assume that we know the coefficients~$\tilde{\alpha}(L,D)$ for all values of~$L$ such that~$L\le L_\textrm{MAX}$, but not for larger values of~$L$ --- in particular, we have~$L_\textrm{MAX}=17$.
In this case, we can exploit the inequality in~(\ref{e:lemma4}) to manipulate the coefficients~$\tilde{\alpha}(L,D)$ for~$L>L_\textrm{MAX}$.
The obtained results are reported in Fig.~\ref{fig:ub1}, for all values of~$D$ in~$\{0,1,\dots,17\}$ and $L_\textrm{MAX}=17$.
The resulting bounds, referred to as~$C_1^*$, are actually larger than the capacity~$C_1$ in~(\ref{e:C_1}), because of the use of~(\ref{e:lemma4}) for~$L>L_\textrm{MAX}$.
Hence, the reported curves can be improved when an inequality tighter than~(\ref{e:lemma4}) is exploited to manipulate the coefficients~$\tilde{\alpha}(L,D)$ for~$L>L_\textrm{MAX}$.
In Fig.~\ref{fig:ub1}, the upper bound proposed in~\cite{DiMiPf07} and the lower bounds proposed in~\cite{DrMi07} and~\cite{DrKi07}, which are the best existing bounds that we are aware of, are also reported for comparison.\footnote{As explained in Section~\ref{s:intro}, the lower bound proposed in~\cite{DrKi07} is adopted when~$d\le0.35$, while the one proposed in~\cite{DrMi07} is adopted when~$d>0.35$.}
We point out that the upper bound~$C_1^*$ improves the upper bound presented in~\cite{DiMiPf07} for a wide range of $d$~values, in particular when~$d>0.35$.

\section{The Second Upper Bound\label{s:ub2}}

\begin{figure}
\centering
\includegraphics[width=12cm]{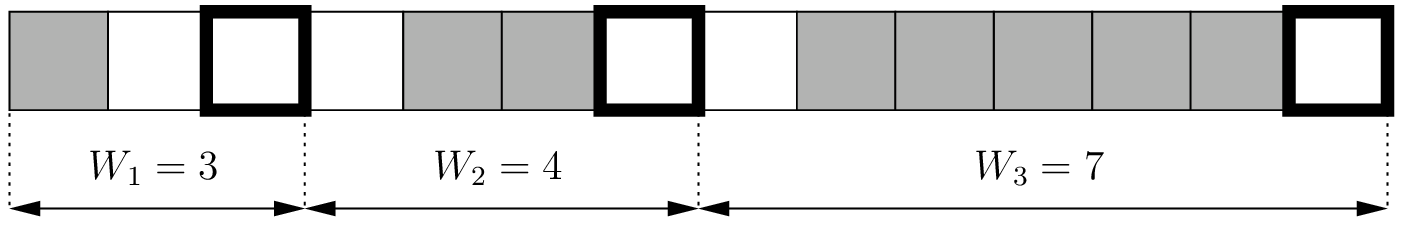}
\caption{A possible realization of the process~$\Wm$, when $R=1$ and~$T=3$. Each white square represents a transmitted bit that is correctly received, while each gray square represents a transmitted bit that is deleted. The positions of the bold-faced bits define the process~$\Wm$.}
\label{fig:W}
\end{figure}

In this section, we derive an upper bound on~$C$ by providing side information on a random process~$\Wm$, defined in the following.
Let $R$ be a non-negative integer parameter and let us assume that the number of bits at the output of the deletion channel is a multiple of~$R+1$, so that $T=M/(R+1)$ is an integer --- this assumption does not affect the capacity evaluation, as in the previous case.
We define $\Wm=\{W_i\}_{i=1}^T$ such that $W_1$ is equal to the position in the transmitted sequence of the \mbox{$[R+1]$-th}~received bit and, for each value of~$i$ in $\{2,3,\dots,T\}$, $W_i$ is equal to the difference between the position in the transmitted sequence of the \mbox{$[(R+1)i]$-th}~received bit and that of the \mbox{$[(R+1)(i-1)]$-th}~received bit.
An example is depicted in Fig.~\ref{fig:W} and discussed in the related caption.
Given the assumption of IID deletions, the process~$\Wm$ is IID too, and each element of~$\Wm$ takes on the value $L+1$ with probability
\begin{equation}
P(W_i=L+1) = (1-d) p(L,R)
\label{e:ub2_pacal}
\end{equation}
according to the Pascal distribution~\cite{Pa91}, for all values of $L$ such that~$L\ge R$.

As in the previous case, an upper bound on the capacity of the deletion channel can be obtained by providing the transmitter and the receiver with genie-aided information on the realizations of~$\Wm$.
We will refer to the capacity per input bit of this genie-aided system as~$C_2$.
Similarly to the previous case, we have $T$~blocks that do not interfere with each other, the $i$-th block having $W_i$ input bits and $R+1$ output bits.
The last input bit of each block can be safely sent uncoded, since both the transmitter and the receiver know that it is correctly received.
Hence, following the same arguments as in the previous section, we get
\begin{eqnarray}
C_2 &=& \lim_{N\to \infty} \frac{1}{N} \sum_{i=1}^T \left[f(W_i-1,R) + 1\right]
\nonumber \\
&=& \frac{1}{E[W_i]} \lim_{T\to \infty} \frac{1}{T} \sum_{i=1}^T \left[f(W_i-1,R) + 1\right]
\nonumber \\
&=& \frac{1}{E[W_i]} E \left[f(W_i-1,R) + 1\right] \;.
\nonumber
\end{eqnarray}
Finally, by exploiting~(\ref{e:ub2_pacal}) and the properties of the Pascal distribution, the upper bound yields
\begin{equation}
C_2 = \frac{(1-d)^2}{R+1} \sum_{L=R}^\infty \left[f(L,R)+1 \right] p(L,R)
\nonumber
\end{equation}
which can be also written as
\begin{eqnarray}
C_2 &=& \underbrace{\frac{(1-d)^2}{R+1} \sum_{L=R}^\infty (R+1) \, p(L,R)}_{1-d} - \frac{(1-d)^2}{R+1} \sum_{L=R}^\infty \left[R-f(L,R)\right] p(L,R)
\nonumber\\
&=& 1-d - \frac{(1-d)^2}{R+1} \sum_{L=R}^\infty \alpha(L,R) p(L,R) \;.
\label{e:C_2}
\end{eqnarray}
Since the coefficients $\alpha(\cdot,\cdot)$ cannot be negative, the bound~(\ref{e:C_2}) is at least as good as the trivial bound $1-d$.
In particular, by combining Lemma~2 with the available outcomes of the BAA, it can be proved that the bound~(\ref{e:C_2}) equals~$1-d$ \mbox{when~$R\in\{0,1\}$}, otherwise it is strictly better.

\begin{figure}
\centering
\includegraphics[width=12cm]{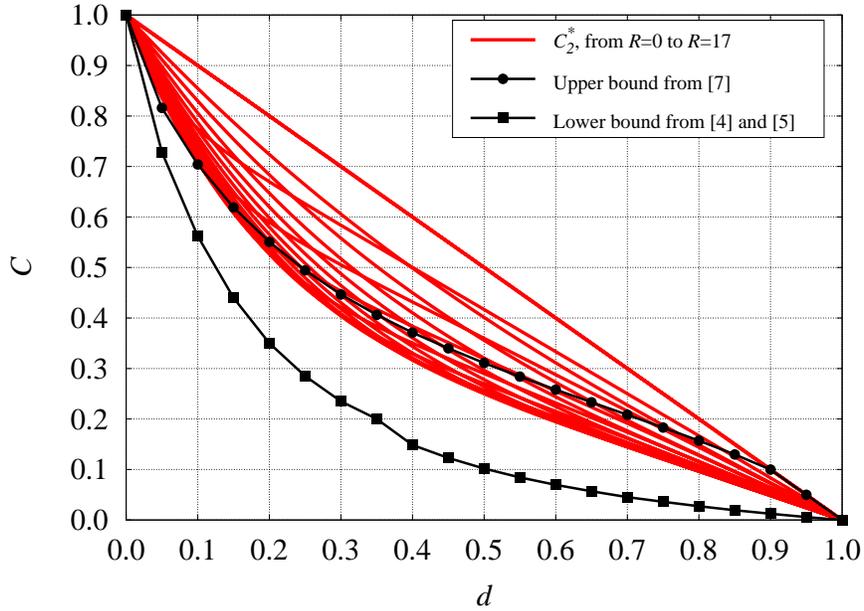}
\caption{Different bounds on the capacity of the deletion channel.}
\label{fig:ub2}
\end{figure}

When~$R>1$, it seems infeasible to evaluate the coefficients $\alpha(L,R)$ for all values of~$L$ required in~(\ref{e:C_2}).
Let us assume that we know the coefficients $\alpha(L,R)$ for all values of~$L$ such that~$L\le L_\textrm{MAX}$, but not for larger values of~$L$ --- in particular, we have~$L_\textrm{MAX}=17$.
In this case, we can exploit~(\ref{e:lemma2}) to manipulate the coefficients~$\alpha(L,R)$ \mbox{for~$L>L_\textrm{MAX}$}, obtaining
\begin{equation}
C_2^* = \frac{(1-d)^2}{R+1} \sum_{L=R}^{L_\textrm{MAX}} \left[\alpha(L_\textrm{MAX},R)-\alpha(L,R)\right] p(L,R) + (1-d) \left[1-\frac{\alpha(L_\textrm{MAX},R)}{R+1}\right]
\label{e:C_2*}
\end{equation}
after a few straightforward manipulations --- the bound is referred to as~$C_2^*$ because it is actually larger than the capacity~$C_2$ in~(\ref{e:C_2}).
The obtained results are reported in Fig.~\ref{fig:ub2}, for all values of~$R$ in~$\{0,1,\dots,17\}$ and $L_\textrm{MAX}=17$.
Clearly, such curves can be improved when an inequality tighter than~(\ref{e:lemma2}) is exploited to manipulate the coefficients~$\alpha(L,R)$ \mbox{for~$L>L_\textrm{MAX}$.}
In Fig.~\ref{fig:ub2}, the upper bound proposed in~\cite{DiMiPf07} and the lower bounds proposed in~\cite{DrMi07} and~\cite{DrKi07} are also reported for comparison.
We point out that the upper bound~$C_2^*$ improves the upper bound presented in~\cite{DiMiPf07} for most values of~$d$, in particular when~$d>0.1$, and, for large values of~$d$, the gap from the best lower bound is now roughly halved.

\section{The Third Upper Bound\label{s:ub3}}

\begin{figure}
\centering
\includegraphics[width=12cm]{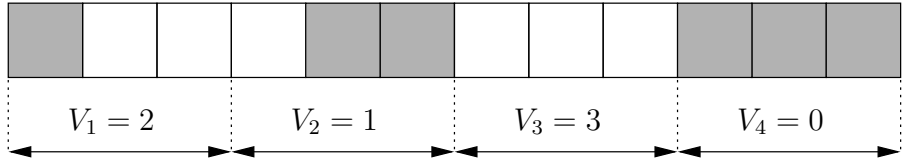}
\caption{A possible realization of the process~$\Vm$, when $L=3$ and~$Q=4$. Each white square represents a transmitted bit that is correctly received, while each gray square represents a transmitted bit that is deleted.}
\label{fig:V}
\end{figure}

In this section, we derive an upper bound on~$C$ by providing side information on a random process~$\Vm$, defined in the following.
Let $L$ be a positive integer parameter, based on which we partition the input sequence~$\Xm$ into subsequences~$\{\Xm_i\}$ of $L$~consecutive bits.
Formally, we define
\begin{equation}
\Xm_i = (X_{(i-1)L+1}, X_{(i-1)L+2}, \dots , X_{iL}), \quad \forall i\ge 1 \;.
\nonumber
\end{equation}
For example, when $L=3$, we have $\Xm_1 = (X_1, X_2, X_3)$, $\Xm_2 = (X_4, X_5, X_6)$, $\Xm_3 = (X_7, X_8, X_9)$, and so on.
We assume that $N$ is a multiple of $L$, and thus that there are exactly $Q=N/L$ subsequences~$\{\Xm_i\}_{i=1}^Q$ --- this assumption does not affect the capacity evaluation, as in the previous cases.
We then partition the output sequence~$\Ym$ into $Q$~subsequences~$\{\Ym_i\}_{i=1}^Q$, where, for each value of $i$ in~$\{1,2,\dots,Q\}$, $\Ym_i$ includes the received bits related to the input subsequence~$\Xm_i$.
Finally, we define the random process~$\Vm = \{V_i\}_{i=1}^Q$ such that, for each value of $i$ in~$\{1,2,\dots,Q\}$, $V_i$ denotes the number of bits in the subsequence~$\Ym_i$.
An example is depicted in Fig.~\ref{fig:V} and discussed in the related caption.
Given the assumption of IID deletions, the process~$\Vm$ is IID too, and each element of~$\Vm$ takes on the value $R$ in~$\{0,1,\dots,L\}$ with probability~$p(L,R)$, according to the binomial distribution.

As in the previous cases, an upper bound on the capacity of the deletion channel can be obtained by providing the transmitter and the receiver with genie-aided information on the realizations of~$\Vm$.
We will refer to the capacity per input bit of this genie-aided system as~$C_3$.
Similarly to the previous cases, we have $Q$~blocks that do not interfere with each other, the $i$-th block having $L$~input bits and $V_i$~output bits.
Hence, using similar arguments as in the previous sections, we get
\begin{eqnarray}
C_3 &=& \lim_{N\to \infty} \frac{1}{N} \sum_{i=1}^Q f(L,V_i)
\nonumber \\
&=& \frac{1}{L} \lim_{Q\to \infty} \frac{1}{Q} \sum_{i=1}^Q f(L,V_i)
\nonumber \\
&=& \frac{1}{L} \sum_{R=0}^L f(L,R) p(L,R)
\nonumber
\end{eqnarray}
which can be also written as
\begin{eqnarray}
C_3 &=& \underbrace{\frac{1}{L} \sum_{R=0}^L R \, p(L,R)}_{1-d} - \frac{1}{L} \sum_{R=0}^L \left[R-f(L,R)\right] p(L,R)
\nonumber\\
&=& 1-d - \frac{1}{L} \sum_{R=0}^{L} \alpha(L,R) p(L,R) \;.
\label{e:C_3}
\end{eqnarray}
Hence, since the coefficients $\alpha(\cdot,\cdot)$ cannot be negative, the bound~(\ref{e:C_3}) is at least as good as the trivial bound $1-d$.
In particular, by combining Lemma~2 with the available outcomes of the BAA, it can be proved that the bound~(\ref{e:C_3}) equals~$1-d$ \mbox{when~$L\in\{1,2\}$}, otherwise it is strictly better.
Note that, unlike the previous cases, the bound~$C_3$ does not involve an infinite series.

\begin{figure}
\centering
\includegraphics[width=12cm]{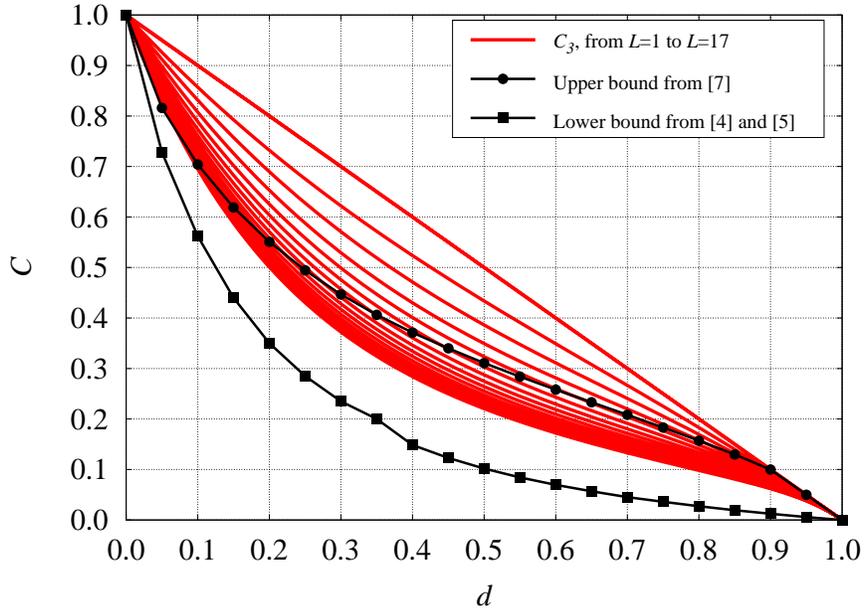}
\caption{Different bounds on the capacity of the deletion channel.}
\label{fig:ub3}
\end{figure}

The upper bound~(\ref{e:C_3}) is plotted in Fig.~\ref{fig:ub3}, together with the upper bound proposed in~\cite{DiMiPf07} and the lower bounds proposed in~\cite{DrMi07} and~\cite{DrKi07}.
For each value of~$L$ for which we could run the BAA, the bound~$C_3$ improves as $L$~increases --- we conjecture that this behavior holds for any value of~$L$ (see Section~\ref{s:discussion}).
Note that the considered approach significantly improves the bound presented in~\cite{DiMiPf07} for most values of the deletion probability~$d$, in particular when~$d>0.08$.


\section{The Fourth Upper Bound\label{s:ub4}}

Given any positive value of the integer parameter~$L$, we can define a system identical to the deletion channel, in which the receiver knows the realizations of the process~$\Vm$ defined in the previous section, while the transmitter does not.
In this case, it is useful to think of the system as if there were a ``parallel'' channel that provides the sequence~$\Vm$ to the receiver.
The capacity per input bit of this system, which will be denoted by~$C_4$, is definitely an upper bound on the capacity~(\ref{e:C}), since, when the parallel output~$\Vm$ is neglected, the original deletion channel is obtained.
Moreover, the upper bound~$C_4$ cannot be larger than~$C_3$ for the same value of~$L$, since the system with capacity~$C_3$ reduces to the system with capacity~$C_4$ when the transmitter neglects the side information on the process~$\Vm$.

\begin{table}
\centering
\begin{tabular}{||c||c|c|c|c|c|c|c||}
\hline
\hline
 & \multicolumn{7}{|c||}{$P(\Ym_i|\Xm_i)$} \\
\hline
$\Xm_i$ & $\Ym_i=\varnothing$ & $\Ym_i=0$ & $\Ym_i=1$ & $\Ym_i=00$ & $\Ym_i=01$ & $\Ym_i=10$ & $\Ym_i=11$ \\
\hline
\hline
$00$ & $d^2$ & $2d(1-d)$ & $0$ & $(1-d)^2$ & $0$ & $0$ & $0$ \\
\hline
$01$ & $d^2$ & $d(1-d)$ & $d(1-d)$ & $0$ & $(1-d)^2$ & $0$ & $0$\\
\hline
$10$ & $d^2$ & $d(1-d)$ & $d(1-d)$ & $0$ & $0$ & $(1-d)^2$ & $0$ \\
\hline
$11$ & $d^2$ & $0$ & $2d(1-d)$ & $0$ & $0$ & $0$ & $(1-d)^2$ \\
\hline
\hline
\end{tabular}
\caption{Transition probabilities for the evaluation of $C_4$ ($L=2$).}
\label{tab:aux_imp_3_2}
\end{table}

As for the system considered in the previous section, we have $Q$~blocks that do not interfere with each other, so that a discrete memoryless channel results.
For each use of this channel, we still have an input sequence of $L$~bits and, with probability~$p(L,R)$, an output sequence of $R$~bits, but now the value of~$R$ is unknown to the transmitter.
Hence, all transmitted sequences must be taken from the same distribution, and no longer from a distribution matched to the number of deletions in the current channel use.
Consequently, the results related to the auxiliary channel introduced in Section~\ref{s:auxiliary} cannot be exploited here.
Formally, we get
\begin{eqnarray}
C_4 &=& \lim_{N\to \infty} \max_{P(\Xm)} \frac{1}{N} I(\Xm;\Ym,\Vm)
\nonumber \\
&=& \frac{1}{L} \lim_{Q\to \infty} \max_{P(\Xm)} \frac{1}{Q} I(\Xm;\Ym,\Vm)
\nonumber \\
&=& \frac{1}{L} \max_{P(\Xm_i)} I(\Xm_i;\Ym_i) \;.
\label{e:C_4}
\end{eqnarray}
When $L=1$, this auxiliary channel reduces to the erasure channel, so that~$C_4=1-d$.
In any other case, we could not find a closed-form expression of~$C_4$, and still resorted to the BAA.
To run the BAA, we need the transition probabilities characterizing the channel, as those reported in Table~\ref{tab:aux_imp_3_2} for the case~$L=2$.
We point out that, unlike the auxiliary channel considered in Section~\ref{s:auxiliary}, the transition probabilities now depend on the value of~$d$, so that the BAA must be run for each value of the deletion probability.

\begin{figure}
\centering
\includegraphics[width=12cm]{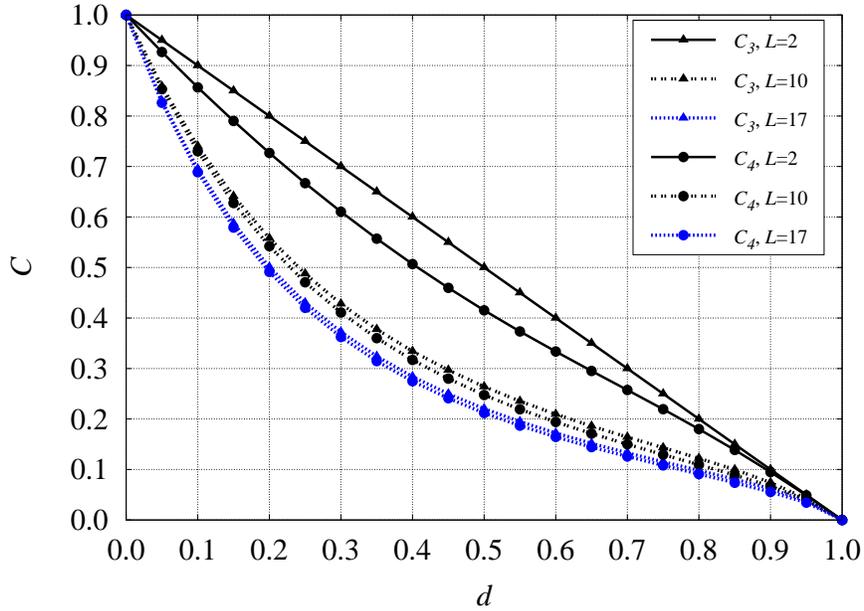}
\caption{Different upper bounds on the capacity of the deletion channel.}
\label{fig:ub4}
\end{figure}

The upper bounds $C_3$ and~$C_4$ are compared in Fig.~\ref{fig:ub4} for three different values of~$L$ --- in both cases, $L=17$ is the largest value for which we could run the BAA.
We point out that the difference between the two bounds, yet $C_4$ is rigorously tighter for each value of~$L$, tends to vanish as~$L$ increases.
This is due to the fact that, for large values of~$L$, the number of deletions for every $L$ transmitted bits is very likely to be close to~$dL$, so that the advantage of knowing the actual number of such deletions (as it happens to the transmitter for the system with capacity~$C_3$) tends to vanish.
As for the bound~$C_3$, for each value of~$L$ for which we could run the BAA, the bound~$C_4$ improves as $L$~increases, and we conjecture that this behavior holds for any value of~$L$ (see Section~\ref{s:discussion}).

\section{Discussions on the Proposed Upper Bounds\label{s:discussion}}

In Table~\ref{tab:upper}, we report a comparison between the best upper bounds found in this paper, that is, $C_4$ with $L=17$ for $d\le0.83$ and $C_2^*$ with $L_\textrm{MAX}=17$ for $d>0.83$, and the existing upper bounds that we are aware of.
We remark that the proposed approaches lead to a new state-of-the-art upper bound on the capacity of the deletion channel for most values of~$d$, as evident from the table (where the best values are shown in bold face).

\begin{table}
\centering
\begin{tabular}{||c||c|c|c||}
\hline
\hline
$d$ & Erasure-channel bound & Bound from~\cite{DiMiPf07} & Proposed bound \\
\hline
\hline
0.01 & 0.990 & not given in~\cite{DiMiPf07} & \bf 0.963 \\
\hline
0.02 & 0.980 & not given in~\cite{DiMiPf07} & \bf 0.926 \\
\hline
0.03 & 0.970 & not given in~\cite{DiMiPf07} & \bf 0.891 \\
\hline
0.04 & 0.960 & not given in~\cite{DiMiPf07} & \bf 0.858 \\
\hline
0.05 & 0.950 & \bf 0.816 & 0.826 \\
\hline
0.10 & 0.900 & 0.704 & \bf 0.689 \\
\hline
0.15 & 0.850 & 0.619 & \bf 0.579 \\
\hline
0.20 & 0.800 & 0.551 & \bf 0.491 \\
\hline
0.25 & 0.750 & 0.494 & \bf 0.420 \\
\hline
0.30 & 0.700 & 0.447 & \bf 0.362 \\
\hline
0.35 & 0.650 & 0.406 & \bf 0.315 \\
\hline
0.40 & 0.600 & 0.371 & \bf 0.275 \\
\hline
0.45 & 0.550 & 0.340 & \bf 0.241 \\
\hline
0.50 & 0.500 & 0.311 & \bf 0.212 \\
\hline
0.55 & 0.450 & 0.284 & \bf 0.187 \\
\hline
0.60 & 0.400 & 0.258 & \bf 0.165 \\
\hline
0.65 & 0.350 & 0.233 & \bf 0.144 \\
\hline
0.70 & 0.300 & 0.208 & \bf 0.126 \\
\hline
0.75 & 0.250 & 0.183 & \bf 0.108 \\
\hline
0.80 & 0.200 & 0.157 & \bf 0.091 \\
\hline
0.85 & 0.150 & 0.130 & \bf 0.073 \\
\hline
0.90 & 0.100 & 0.100 & \bf 0.049 \\
\hline
0.95 & 0.050 & 0.064 & \bf 0.025 \\
\hline
0.96 & 0.040 & not given in~\cite{DiMiPf07} & \bf 0.020 \\
\hline
0.97 & 0.030 & not given in~\cite{DiMiPf07} & \bf 0.015 \\
\hline
0.98 & 0.020 & not given in~\cite{DiMiPf07} & \bf 0.010 \\
\hline
0.99 & 0.010 & not given in~\cite{DiMiPf07} & \bf 0.005 \\
\hline
\hline
\end{tabular}
\caption{Different upper bounds on the capacity of the deletion channel.}
\label{tab:upper}
\end{table}

We believe that the values reported in Table~\ref{tab:upper} could be improved if it were possible to run the BAA for longer sequences.
In particular, our conjecture is formalized in the following.

\vbox{
\textit{Conjecture 1:}
\begin{itemize}
\item the bound~$C_1$ does not worsen as $D$ increases;
\item the bound~$C_2$ does not worsen as $R$ increases;
\item the bound~$C_3$ does not worsen as $L$ increases;
\item the bound~$C_4$ does not worsen as $L$ increases.
\end{itemize}}
These conjectures are based on the amount of genie-aided information, that is, the entropy per input bit of the revealed processes.
The idea is that the lower the entropy per input bit of the revealed information, the tighter the upper bound.
For example, let us consider the bound~$C_1$: if we reveal the position of one deletion every~100, we expect a tighter bound than if we reveal the position of one deletion every~3.
Unfortunately, we could not completely prove the conjectures listed above, but we were able to derive closely related results.
For example, we can prove that $C_4$~does not increase when $L$~is replaced by any positive multiple of~$L$.
It is sufficient to note that, when $L=\ell$, $\Vm$ carries the same information as when $L=n\ell$ ($\forall n>0$), plus some additional information.
Hence, we get
\begin{eqnarray}
\left. \max_{P(\Xm_i)} I(\Xm_i;\Ym_i) \right|_{L=n\ell} \le n \left. \max_{P(\Xm_i)} I(\Xm_i;\Ym_i) \right|_{L=\ell}
\nonumber
\end{eqnarray}
which, according~(\ref{e:C_4}), proves that $C_4$~does not increase when $L=\ell$~is replaced by~$L=n\ell$.

We now discuss the behavior of the proposed upper bounds for limiting values of~$d$, that is, $d\to 0^+$ and~$d\to 1^-$.
In particular, after straightforward manipulations, the following results can be obtained
\begin{eqnarray}
\lim_{d\to 0^+} \frac{1-C_2^*}{d} &=& \alpha(R+1,R) + 1 = \tilde{\alpha}(R+1,1) + 1
\label{e:C_2*_lim0}\\
\lim_{d\to 0^+} \frac{1-C_3}{d} &=& \alpha(L,L-1) + 1 = \tilde{\alpha}(L,1) + 1
\label{e:C_3_lim0}\\
\lim_{d\to 1^-} \frac{C_2^*}{1-d} &=& 1-\frac{\alpha(L_\textrm{MAX},R)}{R+1}
\label{e:C_2*_lim1}
\end{eqnarray}
which are valid for any finite value of $R$, $L$, and~$L_\textrm{MAX}$.
The limits reported above are the only ones leading to closed-form expressions that do not reduce to the trivial erasure-channel bound.

The limit for small values of~$d$ is determined by the coefficient~$\tilde{\alpha}(L,1)$, some values of which are reported in Table~\ref{tab:talpha} --- note that the coefficients in (\ref{e:C_2*_lim0}) and~(\ref{e:C_3_lim0}) are identical, except for the name of the parameters.
The best value that we have found so far is
\begin{equation}
\lim_{d\to 0^+} \frac{1-C_3}{d} = 4.19
\label{eq:C_3_lim0_special}
\end{equation}
obtained when $L=22$.
Other than the erasure-channel bound, we are not aware of any upper bound that leads to closed-form limiting expressions comparable with the reported one.
We believe that~(\ref{eq:C_3_lim0_special}) could be improved if it were possible to run the BAA for longer sequences, as formalized in the following.

\begin{table}
\centering
\begin{tabular}{||c||c|c|c|c|c|c|c|c|c|c|c|c|c||}
\hline
\hline
$L$ & 10 & 11 & 12 & 13 & 14 & 15 & 16 & 17 & 18 & 19 & 20 & 21 & 22\\
\hline
$\tilde{\alpha}(L,1)$ & 2.08 &2.21 & 2.33 & 2.44 & 2.55 & 2.64 & 2.73 & 2.82 & 2.90 & 2.98 & 3.05 & 3.12 & 3.19 \\
\hline
\hline
\end{tabular}
\caption{Coefficient $\tilde{\alpha}(L,1)$.}
\label{tab:talpha}
\end{table}

\textit{Conjecture 2:} For all values of $L$, the following holds
\begin{equation}
\textrm{if} \quad \hat L > L \quad \textrm{then} \quad \tilde{\alpha}(\hat L,1) \ge \tilde{\alpha}(L,1) \;.
\label{e:conj2}
\end{equation}
We wish to prove this conjecture since it would imply that the asymptotic upper bound~(\ref{e:C_3_lim0}) does not worsen as $L$~increases.
Additionally, a strict inequality in~(\ref{e:conj2}), which holds for all available outcomes of the BAA, would imply that the asymptotic upper bound~(\ref{e:C_3_lim0}) improves as $L$~increases.
Lemma~6 gives a partial proof of~(\ref{e:conj2}).
We point out that the limiting value~(\ref{e:C_3_lim0}) may not be limited, since~(\ref{e:lemma8_proof}) does not satisfy any convergence criterion~\cite{Ru74}.

The limit for large values of~$d$ leads to similar considerations.
In particular, the best value that we have found so far is
\begin{equation}
\lim_{d\to 1^-} \frac{C_2^*}{1-d} = 0.49 \;,
\label{e:C_2*_lim1_special}
\end{equation}
obtained by~(\ref{e:C_2*_lim1}) when $R=8$ and~$L_\textrm{MAX}=17$.
Note that, according to~(\ref{e:lemma2}), the reported value could be improved by running the BAA for longer sequences, which unfortunately seems infeasible.
We point out that~(\ref{e:C_2*_lim1_special}) improves the limiting upper bound
\begin{equation}
\lim_{d\to 1^-} \frac{C}{1-d} \le 0.7918
\nonumber
\end{equation}
derived in~\cite{DiMiPf07}, and closes the gap from the limiting lower bound
\begin{equation}
\lim_{d\to 1^-} \frac{C}{1-d} \ge 0.1185
\nonumber
\end{equation}
derived~\cite{MiDr06}.

\section{Two Simple Lower Bounds\label{s:lb}}

In this section, we derive lower bounds on~$C$ by exploiting the random process~$\Vm$ defined in Section~\ref{s:ub3}.
For any input distribution $P(\Xm)$, the following equation holds
\begin{equation}
I(\Xm;\Ym) = I(\Xm;\Ym,\Vm) - I(\Xm;\Vm|\Ym)
\label{e:ami_relation}
\end{equation}
by definition~\cite{CoTh91}.
Moreover, since $I(\Xm;\Vm|\Ym)$ cannot be larger than the entropy $H(\Vm)$ of the process~$\Vm$, we can write
\begin{equation}
I(\Xm;\Ym) \ge I(\Xm;\Ym,\Vm) - H(\Vm)
\label{e:ami_lowerbound}
\end{equation}
from which we get the following lower bound on the capacity of the deletion channel
\begin{equation}
C \ge \lim_{N\to \infty} \frac{1}{N} I(\Xm;\Ym,\Vm)  - \lim_{N\to \infty} \frac{1}{N} H(\Vm)\;.
\label{e:cap_lowerbound}
\end{equation}
If we consider the process $\Vm$ defined before, following the arguments given for the derivation of~(\ref{e:C_4}), we obtain
\begin{eqnarray}
\lim_{N\to \infty} \frac{1}{N} I(\Xm;\Ym,\Vm) &=&  \frac{1}{L} I(\Xm_i;\Ym_i)
\nonumber \\
\lim_{N\to \infty} \frac{1}{N} H(\Vm) &=&  \frac{1}{L} H(V_i) \;,
\nonumber
\end{eqnarray}
so that~(\ref{e:cap_lowerbound}) can be written as
\begin{equation}
C \ge \frac{1}{L} I(\Xm_i;\Ym_i)  + \frac{1}{L} \sum_{R=0}^L p(L,R) \log_2\left[p(L,R)\right] \;.
\label{e:cap_lowerbound_2}
\end{equation}

\begin{figure}
\centering
\includegraphics[width=12cm]{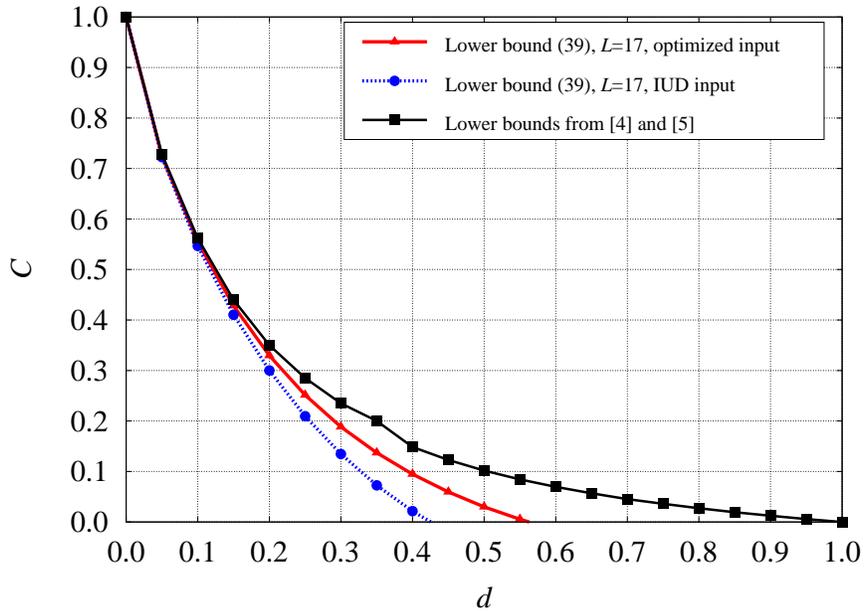}
\caption{Different lower bounds on the capacity of the deletion channel.}
\label{fig:lb}
\end{figure}

In Fig.~\ref{fig:lb}, the lower bound~(\ref{e:cap_lowerbound_2}) is compared with the best lower bound available in the literature, namely the one from~\cite{DrMi07} or the one from~\cite{DrKi07}, depending on the value of~$d$ (see Section~\ref{s:intro}).
For the computation of~(\ref{e:cap_lowerbound_2}), two different input distributions have been considered, that is, the distribution that maximizes $I(\Xm_i;\Ym_i)$, which was considered in the previous section to derive~$C_4$, and IUD input bits.
Note that the difference between the curve related to the optimized input distribution and that related to IUD input bits is not significant for low values of~$d$, which is compliant with the fact that IUD input bits are optimal when~$d=0$.
Interestingly, for low values of~$d$, both distributions lead to a lower bound roughly as good as the reference benchmarks, as evident from Table~\ref{tab:lower} (where the best values are shown in bold face).

\begin{table}
\centering
\begin{tabular}{||c||c|c|c|c||}
\hline
\hline
$d$ & Bound from~\cite{Ga61} & Bound from~\cite{DrKi07} & Bound~(\ref{e:cap_lowerbound_2}), $L=17$, optimized input & Bound~(\ref{e:cap_lowerbound_2}), $L=17$, IUD input \\
\hline
\hline
0.01 & 0.919 & not given in~\cite{DrKi07} & \bf 0.921 & 0.921 \\
\hline
0.02 & 0.858 & not given in~\cite{DrKi07} & \bf 0.862 & 0.862 \\
\hline
0.03 & 0.805 & not given in~\cite{DrKi07} & \bf 0.811 & 0.811 \\
\hline
0.04 & 0.757 & not given in~\cite{DrKi07} & \bf 0.766 & 0.765 \\
\hline
0.05 & 0.713 & \bf 0.728     &     0.724 & 0.722 \\
\hline
0.10 & 0.531 & \bf 0.562     &     0.555 & 0.546 \\
\hline
\hline
\end{tabular}
\caption{Different lower bounds on the capacity of the deletion channel.}
\label{tab:lower}
\end{table}

\section{Conclusions\label{s:conclusions}}

We have presented novel upper bounds on the capacity of the IID binary deletion channel.
All bounds have been obtained by revealing side information on suitable random processes, and by computing the capacity of the resulting genie-aided systems.
The proposed approaches lead to a new state-of-the-art upper bound for most values of the deletion probability~$d$, and provide novel insights on the channel capacity in the limiting scenarios $d\to 0^+$ and~$d\to 1^-$.
As a by-product of our approach, we have also presented simple lower bounds, which turn out not to improve the existing ones.


\def\baselinestretch{1.2}

\end{document}